\documentclass[elsart12,epsf,eqsecnum]{revtex4}
\usepackage{epsfig,epsf}
\def\beq{\begin{equation}}

\def\eeq{\end{equation}}
\begin{document}


%

\def\ap#1#2#3{     {\it Ann. Phys. (NY) }{\bf #1} (19#2) #3}

\def\arnps#1#2#3{  {\it Ann. Rev. Nucl. Part. Sci. }{\bf #1} (19#2) #3}

\def\npb#1#2#3{    {\it Nucl. Phys. }{\bf B#1} (19#2) #3}

\def\plb#1#2#3{    {\it Phys. Lett. }{\bf B#1} (19#2) #3}

\def\prd#1#2#3{    {\it Phys. Rev. }{\bf D#1} (19#2) #3}

\def\prep#1#2#3{   {\it Phys. Rep. }{\bf #1} (19#2) #3}

\def\prl#1#2#3{    {\it Phys. Rev. Lett. }{\bf #1} (19#2) #3}

\def\ptp#1#2#3{    {\it Prog. Theor. Phys. }{\bf #1} (19#2) #3}

\def\rmp#1#2#3{    {\it Rev. Mod. Phys. }{\bf #1} (19#2) #3}

\def\zpc#1#2#3{    {\it Z. Phys. }{\bf C#1} (19#2) #3}

\def\mpla#1#2#3{   {\it Mod. Phys. Lett. }{\bf A#1} (19#2) #3}

\def\nc#1#2#3{     {\it Nuovo Cim. }{\bf #1} (19#2) #3}

\def\yf#1#2#3{     {\it Yad. Fiz. }{\bf #1} (19#2) #3}

\def\sjnp#1#2#3{   {\it Sov. J. Nucl. Phys. }{\bf #1} (19#2) #3}

\def\jetp#1#2#3{   {\it Sov. Phys. }{JETP }{\bf #1} (19#2) #3}

\def\jetpl#1#2#3{  {\it JETP Lett. }{\bf #1} (19#2) #3}


\def\ppsjnp#1#2#3{ {\it (Sov. J. Nucl. Phys. }{\bf #1} (19#2) #3}

\def\ppjetp#1#2#3{ {\it (Sov. Phys. JETP }{\bf #1} (19#2) #3}

\def\ppjetpl#1#2#3{{\it (JETP Lett. }{\bf #1} (19#2) #3} 

\def\zetf#1#2#3{   {\it Zh. ETF }{\bf #1}(19#2) #3}

\def\cmp#1#2#3{    {\it Comm. Math. Phys. }{\bf #1} (19#2) #3}

\def\cpc#1#2#3{    {\it Comp. Phys. Commun. }{\bf #1} (19#2) #3}

\def\dis#1#2{      {\it Dissertation, }{\sf #1 } 19#2}

\def\dip#1#2#3{    {\it Diplomarbeit, }{\sf #1 #2} 19#3 }

\def\ib#1#2#3{     {\it ibid. }{\bf #1} (19#2) #3}

\def\jpg#1#2#3{        {\it J. Phys}. {\bf G#1}#2#3}  

%


\voffset1.5cm


\title{Dense-Dilute Duality at work: dipoles of the target.}
\author{Alex Kovner and  Michael Lublinsky}

\address{Physics Department, University of Connecticut, 2152 Hillside
Road, Storrs, CT 06269-3046, USA}
\date{\today}

\begin{abstract}
We explore the properties of the QCD high energy evolution in the limit of a dilute target. Using the recently established property of selfduality of the evolution operator (hep-ph/0502119), we show how to properly define the target gluon and dipole creation operators in terms of dual Wilson lines (dual eikonal factors). We explain how to expand these operators in terms of the functional derivatives of the color charge density, in the situation when they act on the eikonal factors of the projectile partons. We explicitly calculate the expansion of the high energy evolution operator to fourth order in the functional derivatives. Our result is infrared and ultraviolet finite, but does not coincide with the formula given in hep-ph/0501088. We resolve this discrepancy by showing that the identification of the dipole creation and annihilation operators used in hep-ph/0501088 is incomplete, and provide the required corrections to these definitions. The use of the corrected operators in the calculational framework of hep-ph/0501088 reproduces our result. We also prove that there is no discrepancy between the expansion of the JIMWLK equation and the dualization of the expansion of the weak field limit.
\end{abstract}
\maketitle
\section{Introduction.}
The generalization of the high energy QCD evolution to include Pomeron loops has been a subject of vigorous research in the last year or so \cite{shoshi1},\cite{shoshi2},\cite{ploops},\cite{ll},\cite{kl},\cite{kl1},\cite{something},\cite{genya},\cite{blaizot}. The work \cite{shoshi1} lead to realization that the high density limit of the QCD high energy evolution equation, the so called JIMWLK equation \cite{balitsky},\cite{JIMWLK},\cite{cgc} does not preserve t-channel unitarity and thus the inclusion of Pomeron loops is a must for a fully consistent description of high energy scattering.
This requires the understanding of the evolution operator of the target wave function not only at high density, but in principle at arbitrary density.

The very interesting series of papers \cite{shoshi2},\cite{ploops},\cite{ll} pioneered attempts to derive the low density limit of the evolution, the limit opposite to that in which the JIMWLK equation is valid. This goal has been fully achieved in  \cite{kl}. It was further shown in \cite{something} that the t-channel unitarity together with Lorentz invariance of the scattering matrix require that the evolution operator when properly extended to arbitrary density must be selfdual, see below. It was also shown in \cite{something} that the duality transformation interchanges the high and low density regimes. This Dense-Dilute Duality (DDD) relates the evolution equation derived in \cite{kl} to the JIMWLK equation\cite{footnote1}.

Despite this impressive progress there is still a number of outstanding problems even on the level of comparison of the results of different groups. In particular 
as explained in detail in \cite{kl} the proper account of the noncommutative nature of the charge density operators necessitates the introduction of an extra "ordering" coordinate, which can be identified with the longitudinal coordinate of the target. 
The results of \cite{shoshi2} and \cite{ploops,ll} are presented without invoking 
such an extra coordinate.
Also
the evolution equation of \cite{kl} is written in terms of path ordered exponentials of the functional derivatives with respect to the color charge density, while the final result of \cite{shoshi2} 
(and also of \cite{ploops}) has at most four such functional derivatives. 
In addition, there is some confusion in the literature as to whether the expression for the evolution operator for week target with four derivative term can be obtained via the DDD transformation from the expansion of the JIMWLK equation to the corresponding order. It has been stated in \cite{blaizot} that this cannot be done in apparent contradiction with the proof of \cite{something} of the duality between the full unexpanded expressions.
It is thus important to establish the correspondence between these different approaches and understand their limitations if any. 

The purpose of the present paper is to analyze these questions. Our main results are the following. We show that in the evolution equation derived in \cite{kl} the integrations over the longitudinal coordinate $x^-$ can be performed explicitly as long as one considers the evolution of the products of the charge density operators which are symmetrized with respect to all color indices. Such symmetric combinations arise while averaging the $S$-matrix of the projectile partons in the wave function of a dilute target. We can thus represent the results of \cite{kl} in the same language as \cite{shoshi2} and compare the two expressions directly. The operator of \cite{kl} still contains all powers of the functional derivatives, as opposed to at most four functional derivatives in \cite{shoshi2}. The truncation in number of derivatives corresponds to the assumption that the projectile as well as the target is dilute (the type of assumption used in the dipole model calculations \cite{dipole},\cite{kozlov}). The higher derivative terms in the kernel 
are active if the projectile that impinges on the target in question contains a large number of partons which overlap in transverse plane. The expression of \cite{shoshi2} is derived under assumption that only one projectile dipole can scatter off a given target dipole. From the point of view of the DDD of \cite{something}, this is a restriction forbidding a target dipole to multiply scatter on the projectile. The more complete expression of \cite{kl} does not impose such restrictions and allows for arbitrary number of projectile dipoles to scatter on a given dipole in the target, or by duality allows multiple scatterings of a given target dipole on a projectile.
For a meaningful comparison therefore the more complete expression of \cite{kl} has to be expanded to the four derivative order. Performing this expansion we find that the result of \cite{kl} as expected is directly related by the DDD transformation of \cite{something} to the expansion of the JIMWLK equation. It however does not coincide with the expression given in \cite{shoshi2},\cite{ploops}. 

The root of this discrepancy is in the fact that the expressions for the dipole creation and annihilation operators in terms of the color charge densities and their derivatives used in \cite{shoshi2} are incomplete. The duality transformation allows us to define the target dipole (and gluon)  creation and annihilation operators exactly. In terms of these operators in the large $N_c$ limit, the complete evolution operator of \cite{kl} indeed has the form of a simple $1\rightarrow 2$ dipole creation vertex, which is the starting point of the formulation of \cite{shoshi2}. However expanding these operators in the functional derivatives, we find that they have extra terms compared to the operators postulated in \cite{shoshi2}. The terms absent in \cite{shoshi2} are precisely those which allow the scattering of two projectile dipoles on a single dipole in the target. Although in the kinematics of the dipole model (large but dilute target and projectile) the effect of these terms may be suppressed, they should be kept in the evolution if one wants to keep the "Pomeron splitting vertex" at all away from the dilute dipole model limit\cite{footnote2}. In this regime as well as in the regime when the target has a small number of dipoles these additional terms contribute to the same order in the evolution equation. Restoring these omitted terms in the result of \cite{shoshi2} we find complete agreement with the expansion of the evolution operator of \cite{kl}. We also show that,
contrary to the statement in \cite{blaizot}  there is no discrepancy between the expansion of the evolution equation in the weak field limit  and the dualized expansion of the JIMWLK equation. The two procedures indeed give the same result if both expansions are performed properly.

This paper is organized as follows. In Section 2, after a brief review of the results of \cite{kl} and \cite{something}, we show how to explicitly integrate over the longitudinal coordinate in the evolution equation when the evolution of fully symmetric operators is considered. We then perform the expansion of the evolution operator to fourth order in functional derivatives.
In Section 3 we formulate the dipole model limit of the evolution equation of \cite{kl}
and show that the evolution of the target wave function when expressed in terms of the target dipole degrees of freedom is identical to the dipole limit of the JIMWLK equation \cite{janik},\cite{kl1} viewed from the point of view of the projectile dipoles. We find the expansion of the target dipole creation/annihilation operators in terms of the functional derivative of density, and identify the missing terms in the expressions of \cite{shoshi2}. We also comment on the relation between our formalism and that advocated in \cite{ll}.
Section 4 contains a short discussion of our results.

\section{The Dense-Dilute Duality and the evolution at low density.}
\subsection{Queen's croquet party: the setup.}

The evolution equation for the wave function of a small target $|T\rangle$ has been derived in \cite{kl}. It is conveniently formulated in terms of the evolution of the target weight function $W$ defined so that the expectation value of an arbitrary operator $O[\hat\rho(x)]$ is given by the functional integral
\begin{equation}
\langle T|\,\hat O[\hat\rho(x)]\,|T\rangle\,\,=\,\,\int\, D\rho^{a}\,\, W[\rho(x,x^-)]\,\,O[\rho(x,x^-)]\,.
\label{w}
\end{equation}
Here $\hat\rho^a(x)$ is the operator of the color charge density which depends on the gluonic degrees of freedom in the target.
Note, that although the quantum operator $\hat\rho^a(x)$ depends only on transverse coordinate $x$, the classical variable in the functional integral has an additional coordinate $x^-$. This coordinate has to be introduced since the operators $\hat\rho^a(x)$ at the same transverse position do not commute. The coordinate $x^-$ can be defined to run from $0$ to $1$ without any loss of generality, and we will use this convention in the present paper.
As explained in \cite{kl}, the weight functional $W$ cannot be real as it has to contain a Wess-Zumino term which insures the correct commutation relations of the color charge density operators. Its precise form is not important for our purposes. What is important is that it insures the correct commutation relations between the charge density operators, and also guarantees that the correlation functions of the "classical" fields $\rho^a(x^-_1)...\rho^b(x^-_n)$ do not depend on the exact values of the longitudinal coordinates $x^-_i$, but only on their ordering. We will use this property heavily in the following.

We will also need the definition of the eikonal factor $\alpha(x)$, 
which is defined in general through\cite{JIMWLK,cgc}
\begin{equation}\label{S}
\alpha^a(x,x^-)T^a\,\,=g^2\,\,{1\over \partial^2}(x-y)\,
\left\{S^\dagger(y;0,x^-)\,\,\rho^{a}(y,x^-)\,T^a\,\,S(y;0,x^-)\right\}, \ \ \ \ \ 
S(x;0,x^-)\,\,=\,\,{\cal P}\,\exp\{i\int_{0}^{x^-} dy^-T^a\alpha^a(x,y^-)\}\,.
\end{equation}
where $T^a_{bc}=if^{abc}$ is the generator of the $SU(N)$ group in the adjoint representation.

For a small target this expression can be expanded in powers of $\rho$, and to leading order
\begin{eqnarray}\label{expan}
\alpha^a(x,x^-)T^a\,\,&=&g^2\,\,{1\over \partial^2}(x-y)+...\,
\rho^{a}(y,x^-)\,T^a\,,  \\
S(x;0,x^-)&=&1+i\int_{0}^{x^-} dy^-T^a\alpha^a(x,y^-)-\int_{0}^{x^-}\int_{y^-}^{x^-} dy^-dz^-T^aT^b\alpha^a(x,y^-)\alpha^b(x,z^-)+...\nonumber\\
&=&1+iT^a\int_{0}^{x^-} dy^-\alpha^a(x,y^-)-{1\over 4}\{T^a,T^b\}\int_{0}^{x^-}dy^-\alpha^a(x,y^-)\int_{0}^{x^-} dz^-\alpha^b(x,z^-)+...\nonumber
\end{eqnarray}

The physical meaning of the unitary matrix $S(x;0,1)$ is the scattering matrix of a single gluon projectile at transverse position $x$ on the target $|T\rangle$. The terms in the expansion of the $S$-matrix eq.(\ref{expan}) correspond to one- and two gluon exchanges of the projectile gluon with the target.
In the present paper we restrict our consideration to projectiles containing small number of partons. The eikonal factor for such a projectile can be written in the form analogous to $S(x;0,1)$, see \cite{something}
\begin{equation}
\Sigma^p[\rho]\,\,=\,\,{\cal P}\,\exp\{i\int_{0}^{1} dy^-\int d^2x\,\rho_p^a(x)\,\alpha^a(x,y^-)\}\label{s}
\end{equation}
where the projectile color charge density
\begin{equation}
\rho_p^a(x)\equiv \Sigma_{i}T^a\delta(x-x_i)
\label{tau}
\end{equation}
with $x_i$ - the transverse coordinates of the gluons in the projectile.
Since the projectile charge density $T(x)$ is a sum of delta functions, the effect of the integral over transverse coordinates in eq.(\ref{s}) is simply to symmetrize the product of the $S$-matrices of the individual projectile gluons.

The total $S$-matrix in accordance with eq.(\ref{w}) is given by
\begin{equation}
S^{total}=\int\, D\rho^{a}\,\, W[\rho(x,x^-)]\,\,\Sigma^p[\rho(x,x^-)]\,.
\label{ss}
\end{equation}

If the number of the projectile gluons is not too large (parametrically smaller than $O({1\over g^2})$ and the target is dilute, 
this expression again can be expanded in powers of $\alpha$. The exact expression is of no interest to us here, except for an obvious property which follows from eq.(\ref{expan}), namely that it can be written entirely in terms of sums of products of the integrated
target charge density $\bar\rho(x)\equiv\int_0^1dx^-\rho(x,x^-)$. This will be important in the following.

The $S$ matrix evolves with rapidity as the target wave function changes under boost. The evolution of the wave function in the dilute target limit has been derived in \cite{kl}.
The density correlation functions evolve as \cite{kl}
\begin{eqnarray}
&&{\partial\over\partial Y}\langle\hat\rho^{a_1}(x_1)...\hat\rho^{a_n}(x_n)\rangle=
\frac{\bar\alpha_s}{2\,\pi}\,\int_{x,y,z}\frac{(x\,-\,z)_i}{(x\,-\,z)^2}\,\frac{(y\,-\,z)_i}{(y\,-\,z)^2}
\langle-
{1\over 2}\{\hat\rho^a(x)\rho^a(y),\hat\rho^{a_1}(x_1)...\hat\rho^{a_n}(x_n)\}+ \nonumber \\
&& \hat\rho^b(x)\left[\left(\hat\rho^{a_1}(x_1)+\delta^2(z-x_1)T^{a_1}
\right)...\left(\hat\rho^{a_n}(x_n)+\delta^2(z-x_n)T^{a_n}\right)\right]^{bc}\hat\rho^c(y)\rangle
\label{evolwf}
\end{eqnarray}
In this expression the factors $\hat\rho^{a_i}(x_i)$ are understood as unit matrices in the indices $bc$.

The meaning of this expression is very simple. The target charge density changes due to the emission of an extra gluon in the wave function as a result of boost. The gluon at transverse position $z$ is emitted with probability proportional to the square of the Weizsaker-Williams field
$b_i^a(z)\propto\int d^2x {(x-z)_i\over(x-z)^2}\hat\rho^a(x)$ and contributes an extra adjoint charge to the charge density at this position. The second term in eq.(\ref{evolwf}) is the charge density correlator in the component of the evolved wave function which contains this extra gluon, while the first term is the virtual correction which restores the normalization.

The shift of the charge density in eq.(\ref{evolwf}) can be conveniently  encoded with the help of the finite shift operator 
$e^{T^c\,\frac{\delta}{\delta\hat\rho^c(z)}}$. Thus the evolution of the charge density correlators can be written in a more compact form\cite{kl}
\begin{eqnarray}
\frac{\partial{}}{\partial Y}\,\langle\,\hat \rho^{a_1}\,\cdots\,\hat\rho^{a_n}\rangle &=&
\frac{\bar\alpha_s}{2\,\pi}\,\int_{x,y,z}\frac{(x\,-\,z)_i}{(x\,-\,z)^2}\,\frac{(y\,-\,z)_i}{(y\,-\,z)^2}
\times \nonumber \\
&&\left[\langle-{1\over 2}[\hat\rho^a(x),\,[\hat\rho^a(y),\,\hat\rho^{a_1}\cdots\hat\rho^{a_n}]]\rangle\,\,+\,\,
\langle\hat\rho^a(x)\,\left(\left[e^{T^c\,\frac{\delta}{\delta\hat\rho^c(z)}}\,\,-\,\,1\right]_{ab}
\,\hat\rho^{a_1}\cdots\hat\rho^{a_n}\right)\,\hat\rho^b(y)\rangle\right]
\label{est}
\end{eqnarray}
where the functional derivatives act only on the product of $\rho$'s in parenthesis.
Translating this evolution equation into the evolution equation for the target weight functional
$W$ gives directly \cite{kl}
\begin{equation}
{\partial\over\partial Y}\,W[\rho]\,\,=\,\,\chi^{KLWMIJ}[-i{\delta\over\delta\rho},i\rho]\,\,W[\rho]\,.
\label{dwp}
\end{equation}
with the evolution operator\cite{footnote3}
\begin{equation}\label{KL}
\chi^{KLWMIJ}=-
\frac{\alpha_s}{2\pi^2}\int_{x,y,z} {(z-x)_i(z-y)_i
\over (z-x)^2(z-y)^2}\left\{\rho^a(x,0)\rho^a(y,0)+  \rho^a(x,1)\rho^a(y,1)-2\,  \rho^a(x,0)R(z,0,1)^{ab}\rho^b(y,1)\right\}\,.
\end{equation}
where we have defined
\begin{equation}
R(z;0,x^-)^{ab}=
\left[{\cal P}e^{\int_{0}^{x^-} d z^-T^c{\delta\over\delta\rho^c(z,\,z^-)}}\right]^{ab}
\label{rr}
\end{equation}
Note, that just like in eq.(\ref{est}) the functional derivatives in $R$  act only on $W$ in eq.(\ref{dwp}) and not on the explicit factors $\rho^a(0)$ and $\rho^a(1)$ in Eq. (\ref{KL}).

As we will discuss in more detail in the last section, the exponential factor $R$ can be attributed the meaning of the eikonal factor for the scattering of the extra gluon of the target on the incoming projectile. It is however important to realize that the ordering variable which enters eq.(\ref{rr}) has indeed the meaning of $x^-$ and not $x^+$ as one might naively think.
From the point of view of the derivation given above this is quite obvious. The shift operator $R$ shifts the charge density of the target by the charge of an extra gluon. As a result the incoming partons scatter also on the charge density of this additional gluon. This is most clearly seen if we combine eqs.(\ref{ss}) and (\ref{dwp}) to write the evolution equation for the scattering matrix. Integrating by parts and using the fact that $\chi^{KLWMIJ}$ is hermitian we can write
\begin{equation}
{\partial\over\partial Y}\,S^{total}\,\,=\int\, D\rho^{a}\,\,W[\rho]\,\,
\chi^{KLWMIJ}[-i{\delta\over\delta\rho},i\rho]\,\,\Sigma^p[\rho(x,x^-)]\,\,.
\label{ds}
\end{equation}
where now the functional derivatives in $R$ act on the projectile $S$-matrix.
Since 
\begin{equation}
R(z,0,1)\,\Sigma^p[\rho^a(x,x^-)]\,=\,\Sigma^p[\rho^a(x,x^-)+T^a\delta^2(x-z)]\,\, ,
\end{equation}
we indeed see that the effect of the operator $R$ is to make the projectile scatter off the extra emitted gluon in the target.

We prefer to think of $x^-$ as an ordering variable.
 However if one does insist on interpreting it as a physical longitudinal coordinate in the target wave function, it is also easy to understand qualitatively why the operator $R$ is integrated over $x^-$ and not $x^+$. $R$ represents the scattering of the extra gluon of the target emitted during one step of the evolution. This gluon is emitted
with equal probability at arbitrary longitudinal coordinate $x^-$, and thus the eikonal factor has to be integrated over $x^-$ to take into account all possible longitudinal positions of the emitted gluon. The other longitudinal coordinate $x^+$ does not appear in our formulation at all.
It has the meaning of the ordering variable for the projectile charge density and can be introduced if one wishes to make the projectile charge density $\rho_p^a(x)$ of eq.(\ref{tau}) into a classical commuting variable. We do not have any need for this in the present paper and thus do not pursue this route any further.

Eq. (\ref{KL})  is to be compared with the JIMWLK evolution operator, which describes the evolution of the target wave function when the color charge density is parametrically large ($\rho^a=O(1/\alpha_s)$)
\begin{eqnarray}\label{JIMWLK}
\chi^{JIMWLK}&=&
\frac{\alpha_s}{2\pi^2}\,\int_{x,y,z} {(z-x)_i\,(z-y)_i
\over (z-x)^2\,(z-y)^2} \,\,\times \nonumber \\
&&\left\{ {\delta\over \delta \alpha^a(x,0)}{\delta\over \delta \alpha^a(y,0)}\,+\, 
{\delta\over \delta \alpha^a(x,1)}{\delta\over \delta \alpha^a(y,1)}
\,- \,2\,{\delta\over \delta \alpha^a(x,0)}\,S(z;0,1)^{ab}\,{\delta\over 
\delta \alpha^b(y,1)}\right\}\,.
\end{eqnarray}

The similarity between the two limits of the evolution operator is striking. In fact they are related by a simple duality transformation, which we refer to as the Dense-Dilute Duality\cite{something}
\begin{equation}
\rho\rightarrow -i{\delta\over\delta\alpha}; \ \ \ \ \ \ \ {\delta\over\delta\rho}\rightarrow i\alpha
\label{duality}
\end{equation}
It was proved in \cite{something} that the high energy evolution operator valid at arbitrary charge density, which should incorporate eq.(\ref{KL}) and eq.(\ref{JIMWLK}) as its low and high density limits respectively, must be selfdual, that is it has to  satisfy
\begin{equation}\label{main}
\chi^{HE}[\alpha,\,{\delta\over\delta\alpha}]\,\,=\,\,\chi^{HE}[-i{\delta\over\delta\rho},\,i\rho]\,.
\end{equation}
The DDD transformation eq.(\ref{duality}) interchanges the high and the low density limits, and this is the origin of duality between $\chi^{JIMWLK}$ and $\chi^{KLWMIJ}$.

Although expressions for both $\chi^{JIMWLK}$ and $\chi^{KLWMIJ}$ given above involve the "ordering" variable $x^-$, we know that in the case of JIMWLK this dependence is irrelevant\cite{JIMWLK},\cite{kl1}. In particular, if the JIMWLK evolution operator $\chi^{JIMWLK}$ acts only on functions of the eikonal factor $S(x,0,1)$, it can be rewritten entirely in terms of $S$ and its functional derivatives.
Let us introduce the left and right rotation generators
\begin{equation}\label{LR}
J_R^a(x)=-{\rm tr} \left\{S(x)T^{a}{\delta\over \delta S^\dagger(x)}\right\}, \ \ \ \ J_L^a(x)=
-{\rm tr} \left\{T^{a}S(x){\delta\over \delta S^\dagger(x)}\right\}, \ \ \ \  \ \ \ \ \ \ \ 
J_L^a(x)\,\,=\,\,[S(x)\,J_R(x)]^a,
\end{equation}
The relation between left and right rotations follows from the unitarity of matrix $S$ \cite{kl1}.
We can now write
\begin{equation}
{\delta\over\delta\alpha^a(x,1)}W[S]=-iJ_R^a(x)W[S],\ \ \ \ \ \ \ {\delta\over\delta\alpha^a(x,0)}W[S]=-iJ_L^a(x)W[S]  \label{equiv}
\end{equation}
and the JIMWLK operator becomes
\begin{equation}\label{JIMWLK1}
\chi^{JIMWLK}=-
\frac{\alpha_s}{2\pi^2}\int_{x,y,z} {(z-x)_i(z-y)_i
\over (z-x)^2(z-y)^2}\,\left\{J_L^a(x) \,J_L^a(y)  \,+\, J_R^a(x)\, J_R^a(y)
\,- \,2\,J_L^a(x)\,S(z)^{ab}\,J_R^b(y)\right\}\,.
\end{equation}
Any reference to the ordering variable has disappeared in this equation, it is written entirely in terms of the unitary matrix $S$ and functional derivatives with respect to it.

This leads us to believe that the same should be true for the weak density operator $\chi^{KLWMIJ}$.
The aim of the rest of this section is to substantiate this point and rid ourselves of the ordering variable $x^-$.

In all our calculations from now on we only consider expressions in which the evolution operator
acts on functions of the integrated charge density
\begin{equation}
F[\bar\rho(x)\equiv\int_0^1dx^-\rho(x,x^-)]\label{sym}
\end{equation}
As we have noted above, the $S$-matrix of any projectile on a dilute target is always a function of this type and therefore the set of functions of the type eq.(\ref{sym}) is sufficient for our purposes.

In the operator language eq.(\ref{sym}) means that we are considering only totally symmetrized correlation functions of the charge densities of the type (we omit the transverse coordinates, since the noncommutativity occurs only when the transverse coordinates of the charge density operators coincide)
\begin{equation}
\langle T|\sum_{ \{a_1...a_n\}}\hat\rho^{a_1}...\hat\rho^{a_n}|T\rangle
\label{oper}
\end{equation}
The ordering of the quantum operators in this correlation function obviously 
does not matter, and therefore the relevant correlation function of classical fields will be symmetric under the permutations of the ordering coordinate $x^-$.
\begin{equation}
\langle T|\sum_{\{a_1...a_n\}}\hat\rho^{a_1}...\hat\rho^{a_n}|T\rangle=
\,\,\int\, D\rho^{a}\,\, W[\rho(x,x^-)]\,\,\sum_{\{x^-_1...x^-_n\}}\rho^{a_1}(x^-_1)...\rho^{a_n}(x^-_n)
\end{equation}
Since the correlation functions of the "classical" variables $\rho^a(x^-)$ depend only on the ordering of $x^-_i$'s and not on their values, such correlation function is identically equal to (here our choice of the range of $x^-$ comes in handy, as otherwise we would have to normalize the integrals on the right hand side)
\begin{equation}
\langle T|\sum_{\{a_1...a_n\}}\hat\rho^{a_1}...\hat\rho^{a_n}|T\rangle=
\,\,\int\, D\rho^{a}\,\, W[\rho(x,x^-)]\,\,\int_0^1dx^-_1...\int_0^1dx^-_n\rho^{a_1}(x^-_1)...\rho^{a_n}(x^-_n)
\end{equation}
Thus the correlation functions of the $x^-$ integrals of the "classical" field $\int_0^1 dx^-\rho(x,x^-)$ translate into correlators of fully symmetrized combinations of the quantum operators.

We stress again that this set of functions is sufficient for our present purposes, since scattering amplitude of any projectile on a small target involves explicitly only averages of this type\cite{shoshi2},\cite{ploops}.

Our aim is first, to understand how to get rid of the ordering coordinate $x^-$, and second to expand eq.(\ref{KL}) to fourth order in functional derivatives.
We have two options to proceed, either in terms of the path ordered exponentials as in eq.(\ref{KL}), or directly in terms of the evolution of the correlations of the quantum operators $\hat\rho^a(x)$ as in the initial setup in \cite{kl}. We pursue the first formalism in the body of the paper and give a parallel derivation using the operator formalism in the appendix. The results of the two approaches are identical.

\subsection{Off with her head: getting rid of the ordering variable.}

Our interest is in the averages of the type
\begin{equation}
\int\, D\rho^{a}\,\,W[\rho(x^-)]\,\,\chi^{KLWMIJ}[-i{\delta\over\delta\rho(x^-)},i\rho(x^-)]\,\,F[\bar\rho]\,\,.
\label{dr}
\end{equation}
where the weight functional $W$ can be written as
\begin{equation}
W[\rho(x,x^-)]\,=\,\Sigma[R]\,\delta[\rho(x,x^-)]
\label{ansa}
\end{equation}
The last equation means that the functional Fourier transform of $W[\rho]$ is a function of 
the "eikonal factor" only, namely
\begin{equation}
\int D\rho^ae^{i\int \rho^a(x,t)\alpha^a(x,t)}W[\rho]=\Sigma\left[{\cal P}e^{i\int_{0}^{1} d tT^c\alpha^c(z,\,t)}\right]
\label{ft}
\end{equation}
This must be true for the following reason.
As shown in \cite{something}, the Fourier transform of the weight function $W$ of a state is precisely the $S$-matrix with which this state scatters of some other object. That is to say, if we identify in eq.(\ref{ft}) the function $\alpha$ with the eikonal phase for scattering on some hadron $H$ ($\alpha$ being determined  by the charge density of $H$) , then $\Sigma$ defined via Fourier transform as in eq.(\ref{ft}) is precisely the scattering matrix of our {\it target} state  $|T\rangle$ on this hadronic state $H$. This $S$-matrix clearly must depend only on the scattering matrix of individual gluons and not on
any other properties of the eikonal phase $\alpha(t)$\cite{footnotea}. This is in exact analogy with the statement that in the JIMWLK equation we take $W$ to depend only on the matrix $S$, c.f. discussion around eq.(\ref{LR}). Thus eq.(\ref{ansa}) is merely a statement that the target $|T\rangle$ is a state in the gluonic (or more generally partonic) Hilbert space.
A similar representation for the weight function $W$ was used in \cite{shoshi2} (see however discussion in Section 3). 

With the eqs.(\ref{dr},\ref{ansa}) we can now proceed to express the kernel $\chi^{KLWMIJ}$ in terms of $\bar\rho$ and $\delta/\delta\bar\rho$.

The evolution operator eq.(\ref{KL}) has two structural elements, $R(x;0,1)$ and $\rho(x;0)$ (or alternatively $\rho(x;1)$). The first one is easily dealt with. Consider the operator $R$ acting on 
products of 
\beq\label{tr}
\bar\rho(x)\,=\,\int_0^1dx^-\,\rho(x,\,x^-)
\eeq
Obviously
\beq\label{h5}
\frac{\delta}{\delta\rho^c(x,\,x^-)}\,\bar\rho^a(y)\,=\,
\frac{\delta}{\delta\rho^c(x,\,x^-)}\,\int_0^1dy^-\,\rho^a(y,\,y^-)\,=\,\delta^{ac}\,
\delta^2(x\,-\,y)\,=\,\frac{\delta}{\delta\bar\rho^c(x)}\,\bar\rho^a(y)
\eeq
Thus it is easy to see that
\beq\label{h6}
R(x;0,\,1)\,F[\bar\rho(x)]={\rm exp}\left[T^c\,\frac{\delta}{\delta\bar \rho^c(x)}\right]F[\bar\rho(x)]
\eeq
and more generally
\beq\label{h61}
R(x;0,\,x^-)\,F[\bar\rho(x)]=\,{\rm exp}\left[x^-\,T^c\,\frac{\delta}{\delta
\bar\rho^c(x)}\right]\,F[\bar\rho(x)].
\eeq
We conclude that when acting on functions of $\bar\rho$ the path ordering in $R$ is irrelevant.

Now consider the action of $\rho(x;0)$ and $\rho(x;1)$. It is convenient to think of these operators as acting on the factor $\Sigma[R]$ in eq.(\ref{ansa}) rather than on the function $F$. We can then use explicitly the duality between $\rho$ and $\delta/\delta\alpha$. 

In full analogy with eqs.(\ref{LR},\ref{equiv}) we have the following relation:
\begin{equation}\label{h40}
\rho^a(x,\,0)\,=R^{ab}(x;0,\,1)\,\,\rho^b(x;\,1) 
\end{equation}
Eq.(\ref{h40}) holds as long as both side of it act on an arbitrary function of $R$. It thus holds inside the integral eq.(\ref{dr}). 
The functional derivatives in $R$  act only on the
external factors of charge density (left action), and not on the ones which appear in Eq. (\ref{h40}).
This relation as discussed in the appendix can be also verified in the operator formalism by explicitly commuting the operator $\hat\rho^a(x)$ from left to right (or from right to left) through any  product of operators $\hat\rho^{a_i}(x_i)$ in eq.(\ref{oper}). 
The sum of multiple commutators that arise in this procedure organizes itself into an exponential series which gives eq.(\ref{h40}).

In fact in full analogy with eq.(\ref{equiv}) it is clear that $\rho^a(x;0)$ ($\rho^a(x;x^-)$) is simply the operator of the left (right) rotation when acting on $R(x;0,x^-)$, that is
\begin{equation}
[\rho^a(x;0),R^{bc}(x;0,x^-)]=\left(T^aR(x;0,x^-)\right)^{bc};\ \ \ \ [\rho^a(x;x^-),R^{bc}(x;0,x^-)]=\left(R(x;0,x^-)T^a\right)^{bc}
\end{equation}
Thus the following more general relation (understood in the same operator sense\cite{kl1}), holds for arbitrary $x^-$ 
\begin{equation}\label{h41}
\rho^a(x,\,0)\,=R^{ab}(x;0,\,x^-)\,\,\rho^b(x,\,x^-) 
\end{equation}

Utilizing arbitrarines of $x^-$ in eq.(\ref{h41}) the above equations can be written as
\begin{eqnarray}\label{h4}
\rho^a(x,\,0)\,=\,\int_0^1dx^-\,R^{ab}(x;0,\,x^-)\,\,\rho^b(x,\,x^-)&;&\ \ \ \ \ \ \rho^a(x,\,1)\,=\,\int_0^1dx^-\,R^{\dagger\,ab}(x;x^-,\,1)\,\,\rho^b(x,\,x^-) \nonumber\\
\rho^a(x,\,x^-)\,=\,&\int_0^1&dx_1^-\,R^{ab}(x;x^-,x_1^-)\,\,\rho^b(x,\,x_1^-)
\end{eqnarray}
These relations again hold inside the integral eq.(\ref{dr}). 

From this point onwards we can use the expressions eq.(\ref{h4}) sequentially to express $\rho(x;0)$ in terms of $\bar\rho(x)$.
We start with the expansion
\begin{eqnarray}\label{h7}
\rho^a(x,\,0)&=&\int_0^1dx_1^-\, R^{ab}(x;0,\,x_1^-)\,\rho^b(x,\,x_1^-)\,=\, \\
&&\left[\bar\rho^a(x)\,+\,\int_0^1dx_1^-\,x_1^-\,
\left(T^c\frac{\delta}{\delta\bar\rho^c(x)}\right)_{ab}\,\rho^b(x,\,x_1^-)\,+\,
\int_0^1dx_1^-\,\frac{x_1^{-\,2}}{2}\,
\left(T^c\frac{\delta}{\delta\bar\rho^c(x)}\right)^2_{ab}\,\rho^b(x,\,x_1^-)\right.\nonumber \\
&& +\,\left.
\int_0^1dx_1^-\,\frac{x_1^{-\,3}}{6}\,
\left(T^c\frac{\delta}{\delta\bar\rho^c(x)}\right)^3_{ab}\,\rho^b(x,\,x_1^-)\,+\,\cdots
\right] \nonumber 
\end{eqnarray}
The first term in the right hand side is already expressed directly in terms of  $\bar\rho(x)$.
Now we "unwrap" $\rho(x,\,x_1^-)$ in the rest of the terms using again eq.(\ref{h4}):
\begin{eqnarray}\label{h8}
&&\rho^a(x,\,x_1^-)\,=\,\int_0^1dx_2^-\,R^{ab}(x;x_1^-,\,x_2^-)\,\rho^b(x,\,x_2^-)\,=\, \\
&&\left[\bar\rho^a(x)\,+\,\int_0^1dx_2^-\,(x_2^-\,-x_1^-)\,
\left(T^c\frac{\delta}{\delta\bar\rho^c(x)}\right)_{ab}\,\rho^b(x,\,x_2^-)\,+\,
\int_0^1dx_2^-\,\frac{(x_2^-\,-\,x_1^-)^2}{2}\,
\left(T^c\frac{\delta}{\delta\bar\rho^c(x)}\right)^2_{ab}\,\rho^b(x,\,x_2^-)\right.\nonumber \\
&& +\,\left.
\int_0^1dx_2^-\,\frac{(x_2^-\,-x_1^-)^3}{6}\,
\left(T^c\frac{\delta}{\delta\bar\rho^c(x)}\right)^3_{ab}\,\rho^b(x,\,x_2^-)\,+\,\cdots
\right] \nonumber 
\end{eqnarray}
After substituting this expression into the right hand side of eq.(\ref{h7}) we find that the term with the first functional derivative is expressed entirely in terms of $\bar\rho(x)$. 
To find the second derivative term we continue with expansions of $\rho(x_2^-)$ in the rest of the terms and so on. The result can be written in  the following form
\begin{equation}
\rho^a(x,\,0)\,=\,\left[G(T^c{\delta\over\delta\bar\rho^c(x)})\right]^{ab}\bar\rho^b(x)
\end{equation}
where the function $G$ is defined in terms of the integral of the nested exponentials:
\begin{equation}
G(u)\,=\,\,\int_0^1\,\,
{\rm exp}\left\{x_1^-\,u\,\,\{\,{\rm exp}(x_2^-\,-\,x_1^-)\,u\,\,\{\,{\rm exp}(x_3^-\,-\,x_2^-)\,
u\,...
\}\}\right\} \,\,\Pi_{i=1}^{\infty}\,dx^-_i\,
\end{equation}
Although this expression is somewhat implicit, it's Taylor series expansion is well defined and can be calculated to any given order. Up to fourth order the coefficients of the expansion are given by

\begin{eqnarray}\label{h9}
&&G(u)\,=\,\left[1\,+u
\int_0^1 dx_1^-\,x_1^-\,+u^2
\int_0^1dx_1^-\,dx_2^-\,\left(\frac{x_1^{-\,2}}{2}\,+\,x_1^-(x_2^-\,-\,x_1^-)\right)\right.
\nonumber \\
&&+u^3\,\int_0^1dx_1^-\,dx_2^-\,dx_3^-\,
\left(\frac{x_1^{-\,3}}{6}\,+\,\frac{x_1^{-\,2}}{2}(x_2^-\,-\,x_1^-)\,+\,\frac{x_1^-}{2}
(x_2^-\,-\,x_1^-)^2\,+\,x_1^-(x_2^-\,-\,x_1^-)(x_3^-\,-\,x_2^-)\right)\nonumber \\ 
&&+\,u^4
\int_0^1dx_1^-\,dx_2^-\,dx_3^-\,dx_4^-\,\left(\frac{x_1^{-\,4}}{24} + \frac{x_1^{-\,3}}{6}
(x_2^- -x_1^-) + \frac{x_1^{-\,2}}{2}\left\{\frac{1}{2}(x_2^- -x_1^-)^2
+ (x_2^- -x_1^-)
(x_3^- - x_2^-)\right\} \right. +\nonumber \\
&&\left.\left.
x_1^-\left\{\frac{(x_2^- -x_1^-)^3}{6}+\frac{(x_2^- -x_1^-)^2}{2}(x_3^- - x_2^-)
+\frac{(x_2^- - x_1^-)}{2}(x_3^- - x_2^-)^2 + (x_2^- -x_1^-)(x_3^- -x_2^-)
(x_4^- -x_3^-)\right\} 
\right)\right]
\nonumber\\
&&=1+{1\over 2}u+{1\over 12}u^2-{1\over 720}u^4
\end{eqnarray}
Thus to fourth order in derivatives we have (in fact we will only need the following expressions to the third order, but we give the last term for completeness)
\beq\label{h10}
\rho^a(x,\,0)\,=\,\left[1\,+\,\frac{1}{2}\,\left(T^c\frac{\delta}{\delta\bar\rho^c(x)}\right)\,
+\,\frac{1}{12}\,\left(T^c\frac{\delta}{\delta\bar\rho^c(x)}\right)^2
\,-\,\frac{1}{720}\,\left(T^c\frac{\delta}{\delta\bar\rho^c(x)}\right)^4\right]_{ab}\, 
\bar\rho^b(x)\ \ ,
\eeq
or equivalently
\beq\label{h11}
\rho^a(x,\,0)\,=\,\bar\rho^b(x)\,
\left[1\,-\,\frac{1}{2}\,\left(T^c\frac{\delta}{\delta\bar\rho^c(x)}\right)\,
+\,\frac{1}{12}\,\left(T^c\frac{\delta}{\delta\bar\rho^c(x)}\right)^2
\,-\,\frac{1}{720}\,\left(T^c\frac{\delta}{\delta\bar\rho^c(x)}\right)^4\right]_{ba}\, .
\eeq
The reduction of $\rho(x,\,1)$ is similar.
The first conversion is to $ R(x;1,\,x_1^-)$ rather than $ R(x;0,\,x_1^-)$.
Therefore in the
first expansion of the analog of eq.(\ref{h7}) we have $x_1^-\,\rightarrow\,(x_1^-\,-\,1)$. All the rest of the factors
are the same. As a result the odd terms in the expansion of $\rho(x,\,1)$ have opposite signs to the 
corresponding terms in the expansion of $\rho(x,\,0)$ while the even terms are the same in both expansions
\beq\label{h12}
\rho^a(x,\,1)\,=\,\left[1\,-\,\frac{1}{2}\,\left(T^c\frac{\delta}{\delta\bar\rho^c(x)}\right)\,
+\,\frac{1}{12}\,\left(T^c\frac{\delta}{\delta\bar\rho^c(x)}\right)^2
\,-\,\frac{1}{720}\,\left(T^c\frac{\delta}{\delta\bar\rho^c(x)}\right)^4\right]_{ab}\,
\bar\rho^b(x)
\eeq
Again we stress that all the relations (\ref{h10},\ref{h11},\ref{h12}) are valid only when appearing in expressions of the form eq.(\ref{dr}) with eq.(\ref{ansa}).

\subsection{Who stole the tarts: the four derivative terms.}

With these expressions we can extract the four derivative term of the evolution operator in order to compare it with the result of \cite{shoshi2}.
We rewrite the evolution operator here for convenience 
\beq\label{e1}
\chi^{KLWMIJ}[\rho]\,=-\,\frac{\alpha_s}{2\,\pi^2}
\int_{x,y,z}\,K(x,y,z)\,\left[\rho^a(x,\,0)\,\rho^a(y,\,0)\,+\,
\rho^a(x,\,1)\,\rho^a(y,\,1)\,-\,
2\,\rho^a(x,\,0) R_{ab}(z)\,\rho^b(y,\,1)\right]
\eeq
First we note, that although the kernel in eq.({\ref{e1}) is the Weiszaker-Wiliams kernel, 
\begin{equation}
K(x,y,z)={(x-z)_i(y-z)_i\over(x-z)^2(y-z)^2}
\end{equation}
when acting on a wave function which is globally invariant under the color charge it can be substituted by the dipole kernel
\begin{equation}
M(x,y,z)=-{(x-y)^2\over 2(x-z)^2(y-z)^2}
\end{equation}
The argument here is identical to that given in \cite{oderon}. The kernels $K$ and $M$ differ by terms which do not depend either on $x$ or on $y$. The factors $\rho(x;0)$ and $\rho(x;1)$ in eq.(\ref{e1}) are simply the charge density operators acting directly on the wave function of the target. 
Thus adding to $K$ a term that does not depend on $x$ produces an integral over $x$ of the charge density operator which acts directly on the wave function of the target. If the target state is invariant under the global color charge, this extra term vanishes. A note of caution: the present argument, just like the argument of \cite{oderon} does not allow one to replace $K$ by $M$ for all gauge invariant states\cite{footnote} and therefore in the following we will continue using the Weiszacker-Williams kernel $K$ rather than $M$. However for dipole type states the argument is indeed valid and thus for the purpose of comparison with \cite{shoshi2} the kernels $K$ and $M$ are equivalent.

We can now use the formulae derived in the previous subsection to find the four derivative terms in eq.(\ref{e1}). First, since there is no third derivative contribution in eq.(\ref{h10}), no fourth derivative terms are present in the combination
\beq
\rho^a(x,\,0)\,\rho^a(y,\,0)\,+\,
\rho^a(x,\,1)\,\rho^a(y,\,1)\,-\,
2\,\rho^a(x,\,0)\,\rho^a(y,\,1)
\eeq
In fact this part of the operator contains only two derivatives, and therefore simply reproduces part of the BFKL kernel. All higher derivatives  therefore reside in the expression
\begin{eqnarray}\label{d1}
\rho^a(x,\,0)\,[R(z)\,-\,1]_{ab}\,\rho^b(y,\,1)\,&=&\,
\bar\rho^a(x)\,\left\{
\left[1\,-\,\frac{1}{2}\,\left(T^c\frac{\delta}{\delta\bar\rho^c(x)}\right)\,
+\,\frac{1}{12}\,\left(T^c\frac{\delta}{\delta\bar\rho^c(x)}\right)^2\right]
\right.
\nonumber \\
&&
\left[\left(T^d\frac{\delta}{\delta\bar\rho^d(z)}\right)\,
+\,\frac{1}{2}\,\left(T^d\frac{\delta}{\delta\bar\rho^d(z)}\right)^2
\,
+\,\frac{1}{6}\,\left(T^d\frac{\delta}{\delta\bar\rho^d(z)}\right)^3
\,
+\,\frac{1}{24}\,\left(T^d\frac{\delta}{\delta\bar\rho^d(z)}\right)^4
\right] \nonumber \\
&& \left.
\left[1\,-\,\frac{1}{2}\,\left(T^e\frac{\delta}{\delta\bar\rho^e(y)}\right)\,
+\,\frac{1}{12}\,\left(T^e\frac{\delta}{\delta\bar\rho^e(y)}\right)^2\right]
\right\}_{ab}\bar\rho^b(y)
\end{eqnarray}
Focusing on the four derivative term we find \cite{footnoteb}
\begin{eqnarray}\label{d2}
\chi^{4d}\,=\,\frac{\alpha_s}{24\,\pi^2}
\int_{x,y,z}\,K(x,y,z)\bar\rho^a(x)\,\ \bar\rho^b(y)\,\left\{
\left[T^c\frac{\delta}{\delta\bar\rho^c(z)} \,-\,
T^c\frac{\delta}{\delta\bar\rho^c(x)}\right]\,\,
\left(T^d\frac{\delta}{\delta\bar\rho^d(z)}\right)^2\,\,
\left[ T^e\frac{\delta}{\delta\bar\rho^e(z)}\,-\,
T^e\frac{\delta}{\delta\bar\rho^e(y)}\right]\right. \nonumber \\ \nonumber \\
-\,\,\left(T^f\frac{\delta}{\delta\bar\rho^f(x)}\right)
\left[T^c\frac{\delta}{\delta\bar\rho^c(z)} \,-\,
T^c\frac{\delta}{\delta\bar\rho^c(x)}\right]\,\,
\left(T^d\frac{\delta}{\delta\bar\rho^d(z)}\right)\,\,
\left[ T^e\frac{\delta}{\delta\bar\rho^e(z)}\,-\,
T^e\frac{\delta}{\delta\bar\rho^e(y)}\right] \nonumber \\ \nonumber \\
\left.
-\,\,\left[T^c\frac{\delta}{\delta\bar\rho^c(z)} \,-\,
T^c\frac{\delta}{\delta\bar\rho^c(x)}\right]\,\,
\left(T^d\frac{\delta}{\delta\bar\rho^d(z)}\right)\,\,
\left[ T^e\frac{\delta}{\delta\bar\rho^e(z)}\,-\,
T^e\frac{\delta}{\delta\bar\rho^e(y)}\right]
\left(T^f\frac{\delta}{\delta\bar\rho^f(y)}\right)
\right\}_{ab}
\end{eqnarray}

This operator possesses the required properties of the ultraviolet and infrared finiteness. The ultraviolet finiteness is the fact that if the coordinate $z$ becomes equal to $x$ or $y$, the integrand in eq.(\ref{d2}) vanishes. The infrared finiteness is guaranteed since each individual term in the expression contains at least one derivative with respect to $\rho(z)$. Thus if in the initial wave function there are no color charges beyond some radius $L$, this radius automatically takes on the role of the infrared cutoff in eq.(\ref{d2}).

We can now compare our result with the expression for the four derivative term given in  \cite{shoshi2}
\begin{eqnarray}\label{d211}
\chi_{MSW}^{4d}\,=\,\frac{\alpha_s}{16\,\pi^2\,N_c^2}
\int_{x,y,z}\,K(x,y,z)\,\,\rho^a(x)\,\ \rho^a(y)\,
\left[\frac{\delta}{\delta\rho^b(z)} \,-\,
\frac{\delta}{\delta\ \rho^b(x)}\right]^2
\left[ \frac{\delta}{\delta\rho^c(z)}\,-\,
\frac{\delta}{\delta\rho^c(y)}\right]^2
\end{eqnarray}
The two look quite different. To make the comparison more explicit we have to take the large $N_c$ limit of eq.(\ref{d2}).
This procedure is slightly involved and is discussed in more detail in Appendix B. However to see that the two expression indeed do not coincide, we can for example act with the two kernels on four factors of $\bar\rho$ representing four projectile partons. 
At large $N_c$ the action of $\chi^{4d}$ on a product of two pairs   $\bar\rho^a(x_1)\,\bar\rho^a(x_2) \,\bar\rho^b(x_3)\,\bar\rho^b(x_4)$ is identical to the action of the 
following kernel \cite{thanksal}
\begin{eqnarray}\label{d21}
\chi^{4d}_{N_c}\,=\,\frac{\alpha_s}{48\,\pi^2\, N_c^2}
\int_{x,y,z}\,K(x,y,z)\bar\rho^a(x)\,\ \bar\rho^a(y)\,\ \ \ \ \ \ \ \ \ \ \ \ \ \ \ \ \ \ \ \ \ \ \ \ \ \ \ \ \ \ \ \ \ \ \ \ \ \ \ \ \ \ \ \ \ \ \ \ \ \ \ \ \ \ \ \ \ \ \ \ \ \ \ \ \ \\
\times\left\{2\,
\left[\frac{\delta}{\delta\bar\rho^b(z)} \,-\,
\frac{\delta}{\delta\ \bar\rho^b(x)}\right]\,\,\frac{\delta}{\delta\bar\rho^b(z)}
\left[\frac{\delta}{\delta\bar\rho^c(z)}\,-\,\frac{\delta}{\delta\bar\rho^c(x)}\,-\,\frac{\delta}{\delta\bar\rho^c(y)}\right]\left[\frac{\delta}{\delta\bar\rho^c(z)}-\frac{\delta}{\delta\bar\rho^c(y)}\right]
\right. \nonumber \\
\,+\,\left[\frac{\delta}{\delta\bar\rho^b(z)} \,-\,
\frac{\delta}{\delta\ \bar\rho^b(y)}\right]\,\,\frac{\delta}{\delta\bar\rho^b(z)}
\left[\frac{\delta}{\delta\bar\rho^c(z)}\,\,-2\frac{\delta}{\delta\bar\rho^c(x)}-2\frac{\delta}{\delta\bar\rho^c(y)}\right]\left[\frac{\delta}{\delta\bar\rho^c(z)}-\frac{\delta}{\delta\bar\rho^c(x)}\right]\nonumber\\
\left.
+\,\,\left[\frac{\delta}{\delta\bar\rho^b(z)} \,-\,
\frac{\delta}{\delta\bar\rho^b(x)}\right]\,\,\left[\frac{\delta}{\delta\bar\rho^b(z)}\,-\,
\frac{\delta}{\delta\bar\rho^b(y)}\right]
\frac{\delta}{\delta\bar\rho^c(z)}\,\left[ 2\,\frac{\delta}{\delta\bar\rho^c(z)}\,-\,
\frac{\delta}{\delta\bar\rho^c(x)}\,-\,\frac{\delta}{\delta\bar\rho^c(y)}\right]
\nonumber
\right\}
\end{eqnarray}
This clearly does not coinside with eq.(\ref{d211}).
In the next section we explain the origin of the discrepancy.

We note also that the expansion of our operator eq.(\ref{e1}) yields also three derivative terms, which are not formally subleading in the large $N_c$ limit, the odderon type structure similar to the ones discussed in \cite{ksw},\cite{oderon}. Such terms are also absent in \cite{shoshi2},\cite{ploops}. We will however concentrate on the four derivative terms for definiteness.

\section{Dipoles of the target.}
The Dense-Dilute Duality  between the evolution operator eq.(\ref{e1}) and the JIMWLK operator eq.(\ref{JIMWLK}) allows us to discuss the dipole limit in a way completely analogous to the discussion of the dipole limit in \cite{kl1}.

We remind the reader, that in the framework of the JIMWLK equation one can define a "projectile dipole creation operator" as
\beq\label{dp}
s(x,y)\,\equiv\, {1\over N_c}\,{\rm tr}[S_F(x)\,S_F^\dagger(y)]
\eeq
where $S_F$ is the fundamental representation of the single gluon scattering matrix $S(x;0,1)$ defined in eq.(\ref{S}). The reason one can think of it as a dipole creation operator, is that the presence of an extra dipole in the projectile wave function $|P\rangle$ results in an extra power of $s$ in the expression for the scattering matrix averaged over the projectile wave function
\begin{equation}
\Sigma^p[S]\equiv \langle P|\hat S|P\rangle
\end{equation}
By the same token, applying the functional derivative with respect to $s(x,y)$ to $\Sigma$, diminishes the power of $s$, and thus is identified with the projectile dipole annihilation operator.

It is then very clear how to define creation/annihilation operators for dipoles of the target. In the original derivation of \cite{kl}, the path ordered exponential $R(z)$ appeared directly as a consequence of the presence of an extra emitted gluon in the target wave function at transverse coordinate $z$. Also as we have discussed earlier (see eqs.(\ref{ansa},\ref{ft})) if the {\it target} is scattered on another hadron, each factor of $R$ turns into an eikonal factor of a target gluon. Thus $R(x;0,1)$ should be understood as the creation operator of a target gluon. The target dipole then is created by
\beq\label{u1}
d^\dagger(x,y)=r(x,\,y)\,=\,\frac{1}{N_c}\,{\rm tr}[ R_F(x)\, R_F^\dagger(y)]
\eeq
Conversely, the dipole annihilation operator is simply 
\begin{equation}d(x,y)=\frac{\delta}{\delta r(x,y)}
\label{d}
\end{equation}

To derive the target dipole model we assume in full analogy with \cite{kl1} that the target wave function depends on dipole degrees of freedom only. Referring to our earlier discussion this means that in eq.(\ref{ft}) we take
$\Sigma\,=\,\Sigma[r]\,.$ In this case, we can repeat {\it verbatim} the derivation of the dipole limit given in \cite{kl1}. We recall that in the case of JIMWLK the squares of the left and right rotation operators to leading order in $1/N_c$ are given by \cite{kl1}
\begin{eqnarray}\label{u2}
&&J_R^a(x)J_R^a(y)=J_L^a(x)J_L^a(y)\,=\,\frac{N_c}{2}\,\int d^2u\,d^2v\, 
[\delta(u-y)-\delta(v-y)]\,[\delta(u-x)-\delta(v-x)]\,\,s(u,\,v)\,\,
{\delta \over \delta s(u,\,v)}\nonumber\\
&&J_L^b(x)J_R^a(y)
\,=\,\frac{1}{N_c}\,\int d^2u\,d^2v\,{\rm tr}[S^\dagger(u)\tau^bS(v)\tau^a]\,\,
[\delta(v-y)-\delta(u-y)]\,[\delta(v-x)-\delta(u-x)]\,\,{\delta \over \delta s(u,v)}
\end{eqnarray}

An identical calculation gives in the present case
\beq\label{u3}
\rho^a(x,\,0)\rho^a(y,\,0)\,=\rho^a(x,\,1)\rho^a(y,\,1)=\,\frac{N_c}{2}\,\int d^2u\,d^2v\, 
[\delta(u-y)-\delta(v-y)]\,[\delta(u-x)-\delta(v-x)]\,\,r(u,\,v)\,\,
{\delta \over \delta r(u,\,v)}
\eeq
The evolution operator in the dipole approximation becomes
\begin{eqnarray}\label{dchidip}
 \chi^{dipole}[r]\,=
\,\frac{ \bar{\alpha}_s}{2\,\pi}\,
\int_{x,y,z}\frac{(x-y)^2}{(x-z)^2\,(z\,-\,y)^2}\,\,\left[\,-r(x,\,y)\,+\,
\,r(x, z)\,r(y,z)\,\,\right]
\frac{\delta}{\delta r(x, y)} 
\end{eqnarray}

The first lesson from this calculation is that in terms of the target dipole creation and annihilation operators the evolution is exactly the same as postulated in \cite{shoshi2}. Our evolution operator is simply a $(1\rightarrow 2)$ splitting vertex for the target dipoles.
The second thing to note however, is that the expressions of the dipole creation and annihilation operators used in \cite{shoshi2} are incomplete. In particular, the expressions used in \cite{shoshi2} are
\begin{eqnarray}\label{wimst}
&&d^\dagger_{MSW}(x,y)\,=\,1\,+\,{1\over 4N_c}\left[\frac{\delta}{\delta\bar\rho^a(x)} \,-\,
\frac{\delta}{\delta\bar\rho^a(y)}\right]^2\nonumber\\
&&-\,N_c\,d_{MSW}(x,y)\,=\,\bar\rho^a(x)\,\bar\rho^a(y)\,.
\end{eqnarray} 

On the other hand our creation and annihilation operators are infinite series in powers of functional derivatives of $\bar\rho$. Clearly, any correction to $d^\dagger$ or $d$ of eq.(\ref{wimst}) to order $(\delta/\delta\bar\rho)^4$ and $\bar\rho^2(\delta/\delta\bar\rho)^2$ respectively will affect the form of the evolution operator eq.(\ref{d211}). It is easy to see that such corrections to both, the creation and annihilation operators exist. In particular for the dipole creation operator
eq.(\ref{u1}) yields directly
\begin{eqnarray}\label{relate1}
d^\dagger(x,y)&=&1+{1\over 4N_c}\left[\frac{\delta}{\delta\bar\rho^a(x)} \,-\,
\frac{\delta}{\delta\bar\rho^a(y)}\right]^2+{1\over 32N_c^2}\left[\frac{\delta}{\delta\bar\rho^a(x)} \,-\,
\frac{\delta}{\delta\bar\rho^a(y)}\right]^2\left[\frac{\delta}{\delta\bar\rho^b(x)} \,-\,
\frac{\delta}{\delta\bar\rho^b(y)}\right]^2\nonumber\\
&-&{1\over 96N_c^2}\left[\left(\frac{\delta}{\delta\bar\rho^c(x)}\right)^2\left(\frac{\delta}{\delta\bar\rho^d(x)}\right)^2+\left(\frac{\delta}{\delta\bar\rho^c(y)}\right)^2\left(\frac{\delta}{\delta\bar\rho^d(y)}\right)^2+6\left(\frac{\delta}{\delta\bar\rho^c(x)}\frac{\delta}{\delta\bar\rho^c(y)}\right)^2\right.\nonumber\\
&-&4\left.\left(\frac{\delta}{\delta\bar\rho^a(x)}\right)^2\left(\frac{\delta}{\delta\bar\rho^c(x)}\frac{\delta}{\delta\bar\rho^c(y)}\right)-4\left(\frac{\delta}{\delta\bar\rho^a(y)}\right)^2\left(\frac{\delta}{\delta\bar\rho^c(x)}\frac{\delta}{\delta\bar\rho^c(y)}\right)
\right]\nonumber\\
&+&O\left({\delta\over\delta\bar\rho}^3\right)
\end{eqnarray}
We do not write out explicitly the three derivative terms in $d^\dagger$ as they do not enter $\chi^{4d}$. 
To find the expression of the dipole annihilation operator we compare eq.(\ref{h10}) with
eq.(\ref{u3}). On one hand from eq.(\ref{u3}) we have (as we are only interested in the second derivative terms in $d$, we do not have to keep higher order terms in the expression for $d^\dagger$ in the equations below)
\begin{equation}
\rho^a(x;0)\rho^a(x;0)\,=\,-\,N_c\,r(x,y)\,{\delta\over\delta r(x,y)}\,=\,
-\,N_c\,\left\{1+{1\over 4N_c}\left[\frac{\delta}{\delta\bar\rho^a(x)} \,-\,
\frac{\delta}{\delta\bar\rho^a(y)}\right]^2\right\}\,d(x,y)\ ,
\end{equation}
and on the other hand from eq.(\ref{h10})
\begin{equation}
\rho^a(x;0)\rho^a(y;0)=\left\{1-{1\over 4}\left(T^c\frac{\delta}{\delta\bar\rho^c(x)}\right)
\left( T^a\,\frac{\delta}{\delta\bar\rho^a(y)}\right)+{1\over 12}\left[\left(T^c\frac{\delta}{\delta\bar\rho^c(x)}\right)^2+
\left(T^c\frac{\delta}{\delta\bar\rho^c(y)}\right)^2\right]\right\}_{ab}\bar\rho^a(x)\bar\rho^b(y)\,.
\end{equation} 
On balance we find the following relation
\begin{eqnarray}\label{relate}
-\,N_c\,d(x,y)&=&
\left\{1-{1\over 4N_c}\left[\frac{\delta}{\delta\bar\rho^c(x)}-\frac{\delta}{\delta\bar\rho^c(y)}\right]^2\,-\right. \\
&-&\left.{1\over 4}\left(T^c\frac{\delta}{\delta\bar\rho^c(x)}\right)
\left( T^a\,\frac{\delta}{\delta\bar\rho^a(y)}\right)+{1\over 12}\left[\left(T^c\frac{\delta}{\delta\bar\rho^c(x)}\right)^2+
\left(T^c\frac{\delta}{\delta\bar\rho^c(y)}\right)^2\right]\right\}_{ab}\bar\rho^a(x)\bar\rho^b(y)\,.\nonumber
\end{eqnarray} 

We thus clearly see that the relation between the dipole creation/annihilation operators and the color charge density assumed in \cite{shoshi2} is incomplete. The corrections of $O(\delta^4/\delta\rho^4)$ to $d^\dagger(x,y)$ and $O(\rho^2\delta^2/\delta\rho^2)$ to $d(x,y)$ 
not included in the expressions of \cite{shoshi2},\cite{ploops} change the form of the splitting vertex when 
written in terms of the color charge densities and its functional derivatives. 

As we discuss in the next section, the extra terms in $d$ and $d^\dagger$ not kept in \cite{shoshi2} are responsible for processe when more than one dipoles (gluons) of the projectile scatter on the same dipole (gluon) of the target.
In the regime where both the target and the projectile are dilute but 
large (contain large number of dipoles/gluons but low density thereof) the effects of the extra terms we find on the total scattering amplitude are presumably suppressed. However if the "Pomeron splitting" terms are
important in the regime where the "Pomeron merging", or JIMWLK evolution becomes effective, all four derivative terms should be considered on par.

Finally, it is interesting to establish connection between our approach
and that of Ref. \cite{ll}. In the latter paper the evolution operator has been written in terms of the projectile dipole degrees of freedom  $s$ and $\delta/\delta s$, rather than the target charge density degrees of freedom $\rho$ and $\delta/\delta\rho$. For a dilute target it is easy to relate the two: 
\beq\label{s1}
s(x,\,y)\,=\,1\,-\,\frac{1}{4\,N_c}\,(\alpha^a(x)\,-\,\alpha^a(y))^2
\eeq
Thus 
\beq\label{s2}
\bar\rho^a(x)\,\bar\rho^a(y)\,=\,\,\frac{N_c}{8\,\pi^2\,\alpha_s^2}\,\partial^2_x\,\partial^2_y\,s(x,y) 
\eeq
When acting on $F[s]$ the second derivative with respect to $\rho$ in the large $N_c$ limit 
becomes a single derivative with respect to $s$:  
\beq\label{s3}
\frac{\delta^2\,F[s]}{\delta \bar\rho^a(x)\,\delta \bar\rho^a(y)}\,
=\,\frac{\alpha_s^2\,N_c}{2}\,\int d^2u\,d^2v\, 
[\ln (u-y)^2-\ln (v-y)^2]\,[\ln (u-x)^2-\ln (v-x)^2]\,\,
{\delta \,F[s]\over \delta s(u,\,v)}
\eeq
Similarly, the four derivative term is converted into two derivatives with respect to $s$.
When written in terms of $s$ the four derivative term of \cite{shoshi2} becomes
\beq\label{s4}
\chi^{4d}_{MSW}\,=\,\frac{2\,\bar\alpha_s\,\pi}{\alpha_s^2}\,
 \int_{x,y,z,u,v,r,p} K(x,y,z) \,\gamma(x,z; u,v)\,\gamma(y,z; p,r)\,
\partial^2_x\,\partial^2_y\,s(x,y) \frac{\delta^2} {\delta s(u,v)\,\delta s(p,\,r)}
\eeq
with $\gamma$ standing for the dipole-dipole elastic amplitude and  defined through
\beq\label{s5}
\gamma(x,z;u,v)\,\equiv\,\frac{\alpha_s^2}{32\,\pi^2}\,\ln^2 \left( \frac{(x - 
u)^2\,(z - v)^2}{(x - v)^2\,\,(z - u)^2}\right)
\eeq
After integrating the transverse Laplacians by parts we find that this is equivalent to the $2\rightarrow 1$ merging term of \cite{ll}
(see Eqs. (2.14) and (3.30) of that paper). 
This is not surprising since the underlying physical assumptions
 are the same in both papers.
We thus conclude that the approach of \cite{ll} is limited by the same kinematical range (large but dilute projectile/target) as that of \cite{shoshi2,ploops}.

It was  also argued in \cite{ll} that the merging term 
$ s\,\frac{\delta^2}{\delta s\,\delta s}$ must be accompanied by the virtual contribution of the
form $s\,s\, \frac{\delta^2}{\delta s\,\delta s} $. This term was introduced to preserve the
probabilistic interpretation of the dipole evolution of the projectile wave function(\cite{LL1,LL2}) and is necessary to ensure
the s-channel unitarity: to be consistent
with unitarity any dipole evolution kernel $\chi[s]$ has to vanish both at $s=0$ and $s=1$.
We cannot rule out or confirm the presence of such a term within our present framework, since the present derivation explicitly assumes dilute target. In the dilute target limit the square of the dipole amplitude is linearly related, and thus indistinguishable from the amplitude itself: $s(x,y)s(u,v)=s(x,y)+s(u,v)-1$.

\section{Discussion}
In this paper we continued the study of high energy evolution in the dilute target limit. We have shown how to integrate over the "ordering variable" in the evolution equation in this situation. This allowed us to compare the results of \cite{kl} with those of \cite{shoshi2},\cite{ploops},\cite{ll}. We found that although the expression of \cite{kl} in the dipole model limit reduced precisely to the usual dipole creation vertex, when translated into the language of functional derivatives with respect to $\rho$, our result contains extra terms relative to those in \cite{shoshi2},\cite{ploops}. The origin of these extra terms is the contribution to the target dipole creation/annihilation operators due to scattering of two projectile dipoles on a single dipole of the target.

This last point becomes obvious once one realizes that the meaning of the dual path ordered exponent $R(z)$ is the eikonal phase for scattering of the complete projectile wave function on a single gluon of the target at transverse position $z$ (or equivalently, the eikonal phase of the scattering of one gluon of the target on the projectile). We have already seen this in the discussion in section 2. To make this more explicit, consider for  example a single-gluon projectile impinging on a single-gluon target. The $S$-matrix of one projectile gluon is given by (we neglect the color structure throughout this qualitative discussion)
\begin{equation}\label{ones}
S(x)=1+i\alpha(x)
\end{equation}
The one gluon target is described by the "weight functional"
\begin{equation}
W[\rho]\,=\,R(z)\,\delta(\rho)
\end{equation}
This expression for the weight function is analogous to the one used in \cite{dipole},\cite{shoshi2} for the dipole model. The vacuum is described by a delta function, as an $S$-matrix of any projectile on it is equal to unity.
The $S$-matrix for this process is given by 
\begin{equation}
\int d\rho \,S\,W\,=\,1\,+\,i\,g^2\,{1\over\partial^2}(x,z)
\end{equation}
which is precisely the gluon-gluon scattering amplitude due to a single gluon exchange.
In this toy calculation only the first derivative term in $R$ is active since higher derivatives vanish when acting on the single gluon $S$ matrix eq.(\ref{ones}).
For the projectile which consists of two gluons, the $S$-matrix
is
\begin{equation}
S(x,y)\,=\,(1\,+\,i\,\alpha(x))\,(1\,+\,i\,\alpha(y))
\label{twog}
\end{equation}
The scattering amplitude on a single gluon of a target is then
\begin{equation}
\int d\rho \,S(x,y)\,W\,=\,
1\,+\,i(g^2{1\over\partial^2}(x,z)\,+\,g^2{1\over\partial^2}(y,z))\,-\,
{g^4\over 2}{1\over\partial^2}(x,z){1\over\partial^2}(y,z)
\end{equation}
The quadratic term comes from the action of the second derivative term in $R$ on the quadratic term in $\alpha$ in eq.(\ref{twog}).
Clearly when we repeat this gedankencalculation for the $S$ matrix of $n$ incoming gluons we find that the $n$-derivative term in the expansion of $R$ becomes active, and gives the amplitude for simultaneous scattering of all $n$ incoming gluons on the single gluon of the target. It is straightforward to repeat this argument for the target dipole operator $r$. The four derivative term in the expansion of $r(x,y)$ contributes to the amplitude of the simultaneous scattering of two projectile dipoles on the single dipole of the target. These are precisely the terms that distinguish our expression for $d^\dagger(x,y)$ from the one used in \cite{shoshi2},\cite{ploops},\cite{ll}.

Finally we note that the derivation presented in this paper answers the query raised in \cite{blaizot} about the apparent difference between the dual form of the expansion of $\chi^{JIMWLK}$ and the four derivative term in the weak field limit of the evolution. When expanding the JIMWLK evolution operator one has not only to handle carefully the 
dependence of the eikonal factor $S$ on the ordering variable $x^-$, but also relate $\delta/\delta \alpha(x,1)$ and $\delta/\delta \alpha(x,0)$ to $\delta/\delta\int dx^- \alpha(x,x^-)$.
Expanding  naively only the exponental factors $S(x), \ etc.$ in the JIMWLK kernel 
gives a result which depends on the exact form of the kernel used (i.e. whether it is expressed in terms of $\delta/\delta \alpha(x,1)$ or $\delta/\delta \alpha(x,0)$ or in a mixed representation). The simple reason is that the remaining factors ($\delta/\delta \alpha(x,1)$ or $\delta/\delta \alpha(x,0)$) themselves contain corrections in powers of $\alpha$ when expressed in terms of $\delta/\delta\int dx^- \alpha(x,x^-)$. The proper way to perform the expansion  is by employing the dual variant of the procedure described in Section 2 of the present paper. This in particular generates a term involving the product of functional derivatives in three different spatial positions. 
   The result of the correct expansion is the exact dual of our expressions eqs.(\ref{d2};\ref{d21}).

\appendix
\section{Operator Ordering Approach}
In this appendix we perform the calculations of Section 2 in the language of quantum operator ordering.  
Let us denote a fully symmetrized (with respect to color indices) 
product of $n$ charge density operators $\hat\rho^{a_i}$ by $\{\hat \rho^{a_1}\,\cdots\,\hat\rho^{a_n}\}$. Once the product is fully symmetric, the ordering of the operators is irrelevant. This particular set of correlation functions has therefore a representation in terms of a weight function $W[\rho(x)]$ of a classical field $\rho(x)$, where the field $\rho(x)$ does not have a dependence on an additional ordering variable.
\beq\label{r1}
\frac{1}{n!}\,\langle\{\hat\rho^{a_1}\,\cdots\,\hat\rho^{a_n}\}\rangle\,=\langle
\langle\rho^{a_1}\,\cdots\,\rho^{a_n}\rangle\rangle\ \equiv \,\int\,d\rho^a\,\rho^{a_1}\,\cdots\,
\rho^{a_n}\,\,W[\rho(x)]
\eeq 

The rapidity evolution of this correlation function follows directly from eq.(\ref{est}) which is valid for any product of the charge density operators, including fully symmmetric combinations
\begin{eqnarray}\label{e11}
\frac{\partial{}}{\partial Y}\,\langle\,\{\hat \rho^{a_1}\,\cdots\,\hat\rho^{a_n}\}\rangle &=&
\frac{\bar\alpha_s}{2\,\pi}\,\int_{x,y,z}\frac{(x\,-\,z)_i}{(x\,-\,z)^2}\,\frac{(x\,-\,z)_i}{(x\,-\,z)^2}
\times \\
&&\left[-{1\over 2}\langle[\hat\rho^a(x),\,[\hat\rho^a(y),\,\{\hat\rho^{a_1}\cdots\hat\rho^{a_n}\}]]\rangle\,\,+\,\,
\langle\hat\rho^a(x)\,\left[e^{T^c\,\frac{\delta}{\delta\hat\rho^c(z)}}\,\,-\,\,1\right]_{ab}
\,\{\hat\rho^{a_1}\cdots\hat\rho^{a_n}\}\,\hat\rho^b(y)\rangle\right]\nonumber
\end{eqnarray}
The correlator in the right hand side of eq.(\ref{e11}) 
is not yet given in terms of a fully symmetrized products: the factors
$\hat\rho^a(x)$ and $\hat\rho^b(y)$ are not symmetrized with other operators. Our goal is  to reformulate
 (\ref{e11}) as an equation for the distribution $W$. In order to achieve this we need to rewrite 
(\ref{e11}) in terms  of fully symmetrized combinations of $\hat\rho^{a_i}$. This can be done 
using the canonical commutation relations of operators $\hat\rho$:
\beq\label{r2}
[\hat\rho^a(x),\,\hat\rho^b(y)]\,=\,T_{ab}^c\,\hat\rho^c(x)\,\delta(x\,-\,y)
\eeq

In order to illustrate the operator approach let us first rederive the relation (\ref{h40}). 
Consider a string of $n$ operators $\hat\rho^{a_i}$ multiplied from the left
  by the operator $\hat\rho^b(y)$
$$
\hat\rho^b(y)\,\,\,\hat\rho^{a_1}(x_1)\,\cdots\,\hat\rho^{a_n}(x_n) 
$$
When such a correlation function is represented in terms of a classical field $\rho(x,x^-)$, as in Section 2, the operator $\hat\rho^b(y)$ becomes 
$\rho^b(y,x^-=0) $ as its position is to the left of all the other operators in the correlation function. In order to obtain $\rho^b(y,x^-=1)$ we need to commute the operator $\hat\rho^b(y)$ to the right of the string. 
To see how the exponential operator arises in this procedure we first rewrite the commutation relation (\ref{r2}) in the following form
\beq\label{r20}
\hat\rho^b(y)\,\hat\rho^a(x)\,=\,\hat\rho^a(x)\,\hat\rho^b(y)\,+\,\hat\rho^c(y)T_{cb}^d\,
\frac{\delta}{\delta\hat\rho^d(y)}\,\hat\rho^a(x)\,=\,\hat\rho^a(x)\,\left[1\,+\,
T^d\frac{\delta}{\delta\hat\rho^d(y)}\right]_{bc}\,\hat\rho^c(y)
\eeq
In the last equality in (\ref{r20}), the derivative operator acts  to the left. Applying the procedure twice we obtain:
\beq\label{r21}
\hat\rho^b(y)\,\,\,\hat\rho^{a_1}(x_1)\,\hat\rho^{a_2}(x_2)\,=\,
\hat\rho^{a_1}(x_1)\,\hat\rho^{a_2}(x_2)\,\,\,\left[1\,+\,
T^d\frac{\delta}{\delta\hat\rho^d(y)}\,+\,\frac{1}{2}\,\left(T^d\frac{\delta}{\delta\hat\rho^d(y)}\right)^2
\right]_{bc}\,\hat\rho^c(y)
\eeq
After repeatedly applying the procedure $n$ times, the operator $\hat\rho^b(y)$ is moved to the very right
of the string:
\beq\label{r22}
\hat\rho^b(y)\,\,\hat\rho^{a_1}(x_1)\,\cdots\,\hat\rho^{a_n}(x_n) \,=\,
\hat\rho^{a_1}(x_1)\,\cdots\,\hat\rho^{a_n}(x_n)\,\,
\left[e^{T^c\,\frac{\delta}{\delta\hat\rho^c(y)}}\right]_{bc}\,\hat\rho^c(y)
\eeq
This can be written as the following  relation
\beq\label{r23}
\hat\rho_L^b(y)\,=\,
\left[e^{T^c\,\frac{\delta}{\delta\hat\rho^c(y)}}\right]_{bc}\,\hat\rho_R^c(y)
\eeq
which should be understood in the precise sense defined by (\ref{r22}).
Once the correlation function is represented in terms of classical fields $\rho(x,x^-)$, Eq. (\ref{r23}) becomes the relation
(\ref{h40}). Note that by construction, the functional derivatives $\frac{\delta}{\delta\hat\rho^c(y)}$ in the exponential factor
do not act on its ``parent'' $\hat\rho_R^c(y)$.

We now return to eq.(\ref{e11}) with the aim to rewrite the right hand side of the evolution equation explicitly in terms of the correlation functions of fully symmetrized combinations of $\hat\rho^a$"s. The symmetrization procedure is only nontrivial when the charge density operators are at the same transverse position. Otherwise the operators commute and can be moved through each other without obstruction. We will therefore in the following omit the transverse coordinate and assume that all the operators are at the point $x$ or $y$, depending on the coordinate of the external operator which comes from the evolution kernel.

We begin with the double commutator term in eq.(\ref{e11}). 
Using the property:
$$[A,\,B\,C\,D\,\ldots]\,=\,[A,\,B]\,C\,D\,\ldots\,+\,B\,[A,\,C]\,D\,\ldots\,+\,B\,C\,[A,\,B]\,\cdots\,+\,\cdots
$$
we find
\beq\label{r3}
\left[\hat\rho^a(x),\,\{\hat\rho^{a_1}\cdots\hat\rho^{a_n}\}\right]\,\,=\,\,\sum_i\,T_{a\,a_i}^b\,
\{\hat\rho^{b}\,\hat\rho^{a_1}\,\cdots\,(i)\,\cdots\,\hat\rho^{a_n}\}\,,
\eeq
where $(i)$ denotes the absence of the $i$-th term. The right hand side of (\ref{r3}) is now written
in terms of fully symmetric products.
We can write this as 
\beq\label{r4}
\langle[\hat\rho^a(x),\,\{\hat\rho^{a_1}\cdots\hat\rho^{a_n}\}]\rangle\,=\,\sum_i\,T_{a\,a_i}^b\,
\langle\,\{\hat\rho^b\,\hat\rho^{a_1}\,\cdots\,(i)\,\cdots\,\hat\rho^{a_n}\}\rangle\,=\,T_{a\,b}^c\,
\langle\langle\,\rho^c\,\frac{\delta}{\delta\rho^{b}}\,\left(\rho^{a_1}\,\cdots\,\rho^{a_n}\right)\rangle\rangle
\eeq
Using the same procedure twice we find for the double commutator 
\beq\label{r5}
\langle[\hat\rho^a(x),\,[\hat\rho^a(y),\,\{\hat\rho^{a_1}\cdots\hat\rho^{a_n}\}]]\rangle\,=\,T^c_{ab}\,T_{ae}^d\,
\langle\langle\, \rho^c(x)\,\frac{\delta}{\delta\rho^{b}(x)}\, \rho^d(y)\,\frac{\delta}{\delta \rho^{e}(y)}\,
\rho^{a_1}\,\cdots\,\rho^{a_n}\}\rangle\rangle
\eeq
Thus the double commutator term gives rise to two derivatives with respect to $\rho$ and no higher derivative terms,
for any correlator independently of the number of inserted operators $n$. These are the two derivatives which appear in the BFKL
limit of Eq. (\ref{e11}).

We turn now to the symmetrization of the last term in eq.(\ref{e11}). We are seeking the result in the form of an evolution operator, which can be formulated as a polynomial in derivatives $\delta/\delta\rho$
which act on a symmetrized product of operators $\hat\rho^{a_i}$. Our aim is to calculate the terms up to fourth derivative. 
Since at least one derivative $\delta/\delta\rho(z)$ comes from the expansion of the exponent in eq.(\ref{e11}), we only need to extract terms up to the third derivatives that arise due to the symmetrization procedure.
To extract the terms with up to three derivatives it is sufficient to consider the evolution of the one, two and three point functions. 

We consider $n=1$, $n=2$ and $n=3$ in turn.

$\bullet \ \ n\,=\,1$:
\beq\label{r6}
\hat\rho^a\,\hat\rho^{a_1}\,=\,\frac{1}2\,\{\hat\rho^a,\,\hat\rho^a_1\}\,\,+\,\,T_{a\,a_1}^c\,\hat\rho^c
\eeq

$\bullet \ \ n\,=\,2$:
\beq\label{r7}
\hat\rho^a\,\frac{1}{2}\{\hat\rho^{a_1}\,\hat\rho^{a_2}\}\,=\,\frac{1}{6}\{\hat\rho^a\,\hat\rho^{a_1}\,
\hat\rho^{a_2}\}
\,\,+\,\,\frac{1}{2}\,\left[ \frac{1}{2}\,T_{aa_1}^c\,\{\hat\rho^{a_2}\,\hat\rho^c\}  \,\,+\,\,
 \frac{1}{2}\,T_{aa_2}^c\,\{\hat\rho^{a_1}\,\hat\rho^c\}\right]  \,\,+\,\,
 \frac{1}{12}\,\{T^{a_1}T^{a_2}\}_{ac}\,\hat\rho^c     
\eeq

$\bullet \ \ n\,=\,3$:
\begin{eqnarray}\label{r8}
&&\hat\rho^a\,\frac{1}{6}\{\hat\rho^{a_1}\,\hat\rho^{a_2}\,\hat\rho^{a_3}\}=\frac{1}{24}\{\hat\rho^a\,\hat\rho^{a_1}\,
\hat\rho^{a_2}\,
\hat\rho^{a_3}\}
\,\,+\,\,\frac{1}{2}\,\left[ \frac{1}{6}\,T_{aa_1}^c\,\{\hat\rho^{a_2}\,\hat\rho^c\,\hat\rho^{a_3}\}  \,\,+\,\,
 \frac{1}{6}\,T_{aa_2}^c\,\{\hat\rho^{a_1}\,\hat\rho^c\,\hat\rho^{a_3}\}  \,\,+\,\,
\frac{1}{6}\,T_{aa_3}^c\,\{\hat\rho^{a_2}\,\hat\rho^c\,\hat\rho^{a_3}\} 
   \right]  \nonumber  \\
&&+
\frac{1}{12}\,\left[\frac{1}{2}\{T^{a_1}T^{a_2}\}_{ac}\,\{\hat\rho^c \, \hat\rho^{a_3}\}\,\,+\,\,
\frac{1}{2}\{T^{a_1}T^{a_3}\}_{ac}\,\{\hat\rho^c \, \hat\rho^{a_2}\}\,\,+\,\,
\frac{1}{2}\{T^{a_2}T^{a_3}\}_{ac}\,\{\hat\rho^c \, \hat\rho^{a_1}\}
\right]\,\,
\end{eqnarray}
We immediately infer from eqs.(\ref{r6}-\ref{r8}) that up to (and including) the third derivative term we have
\beq\label{r9}
\langle\hat\rho^a(x)\,\{\hat\rho^{a_1}\cdots\hat\rho^{a_n}\}\rangle\,=\,
\left[1\,+\,\frac{1}{2}\,\left(T^b\,
\frac{\delta}{\delta\rho^b(x)}\right)\,+\,\frac{1}{12}\,\left(T^b\,
\frac{\delta}{\delta\rho^b(x)}\right)^2\right]\,\langle\langle\rho^c(y)\rho^{a_1}
\cdots\rho^{a_n}\rangle\rangle
\eeq
and similarly
\beq\label{r10}
\langle\{\hat\rho^{a_1}\cdots\hat\rho^{a_n}\}\,\hat\rho^b(y)\rangle\,=\,\,\left[1\,-\,\frac{1}{2}\,\left(T^a\,
\frac{\delta}{\delta\rho^a(y)}\right)\,+\,\frac{1}{12}\,\left(T^a\,
\frac{\delta}{\delta\rho^a(y)}\right)^2
\,
\right]_{cb}\,\langle\langle\rho^c(y)\rho^{a_1}\cdots\rho^{a_n}\rangle\rangle
\eeq
This coincides with the results obtained in Section 2 with the help of reordering of the path ordered exponentials.

We have also checked that the operator reordering calculation reproduces the coefficient of the four derivative term given in Section 2. The calculation of the four derivative term proceeds along the same lines as above, but the algebra is significantly lengthier. As it does not add new understanding we are not reproducing it here in any detail.
This concludes the explicit demonstration that the operator symmetrization method is equivalent to reordering of path ordered exponentials.

\section{Large $N_c$ limit.}
In this appendix we discuss the large $N_c$ limit of the kernel $\chi^{KLWMIJ}$. 
This is peripheral to the subject of this paper, but may nevertheless turn out to be useful in future. 
As before we concentrate on the four derivative term, but the bulk of the discussion applies to any term in the derivative expansion of the kernel. For convenience we rewrite the four derivative term
\begin{eqnarray}\label{b1}
\chi^{4d}\,=\,\frac{\alpha_s}{24\,\pi^2}
\int_{x,y,z}\,K(x,y,z)\bar\rho^a(x)\,\ \bar\rho^b(y)\,\left\{
\left[T^c\frac{\delta}{\delta\bar\rho^c(z)} \,-\,
T^c\frac{\delta}{\delta\bar\rho^c(x)}\right]\,\,
\left(T^d\frac{\delta}{\delta\bar\rho^d(z)}\right)^2\,\,
\left[ T^e\frac{\delta}{\delta\bar\rho^e(z)}\,-\,
T^e\frac{\delta}{\delta\bar\rho^e(y)}\right]\right. \nonumber \\ \nonumber \\
-\,\,\left(T^f\frac{\delta}{\delta\bar\rho^f(x)}\right)
\left[T^c\frac{\delta}{\delta\bar\rho^c(z)} \,-\,
T^c\frac{\delta}{\delta\bar\rho^c(x)}\right]\,\,
\left(T^d\frac{\delta}{\delta\bar\rho^d(z)}\right)\,\,
\left[ T^e\frac{\delta}{\delta\bar\rho^e(z)}\,-\,
T^e\frac{\delta}{\delta\bar\rho^e(y)}\right] \nonumber \\ \nonumber \\
\left.
-\,\,\left[T^c\frac{\delta}{\delta\bar\rho^c(z)} \,-\,
T^c\frac{\delta}{\delta\bar\rho^c(x)}\right]\,\,
\left(T^d\frac{\delta}{\delta\bar\rho^d(z)}\right)\,\,
\left[ T^e\frac{\delta}{\delta\bar\rho^e(z)}\,-\,
T^e\frac{\delta}{\delta\bar\rho^e(y)}\right]
\left(T^f\frac{\delta}{\delta\bar\rho^f(y)}\right)
\right\}_{ab}
\end{eqnarray}

Since the kernel does not appear on its own but rather in the integral 
\begin{equation}
\int\, D\rho^{a}\,\,W[\rho]\,\,\chi^{4d}[-i{\delta\over\delta\rho},i\rho]\,\,F([\bar\rho])\,\,.
\label{b2}
\end{equation}
we should be considering the large $N_c$ limit of this expression.
For simplicity we assume that in the expansion of $F$ the charge densities are always pairwise contracted in the colour indices, that the only terms present are of the type
\begin{equation}
\left[\bar\rho^{a_1}(x_1)\bar\rho^{a_1}(y_1)\right]\left[\bar\rho^{a_2}(x_2)\bar\rho^{a_2}(y_2)\right]...\left[\bar\rho^{a_n}(x_n)\bar\rho^{a_n}(y_n)\right]
\end{equation}
This assumption is equivalent to the two gluon exchange approximation for the scattering of each individual dipole of the projectile. 

The simplest $F$ of this form which is not annihilated by the four derivatives in $\chi$ is
\begin{equation}
\left[\bar\rho^{a_1}(x_1)\bar\rho^{a_1}(y_1)\right]\left[\bar\rho^{a_2}(x_2)\bar\rho^{a_2}(y_2)\right]
\end{equation}
It can be checked by explicit calculation that the action of $\chi^{4d}$ on this product of the two pairs is equivalent to the action of $\chi^{4d}_{N_c}$ of eq.(\ref{d21}). However some thought makes it evident that this expression does not produce all the leading large $N_c$ contributions in eq.(\ref{b2}) for more general $F$. The functional $\Sigma$ itself also contains functional derivatives.
In the simplest term where the number of factors of $\bar\rho$ in $F$ is the same as the number of the functional derivatives in $\chi^{4d}$, any functional derivative from $\Sigma$ is forced to act on $\bar\rho$ in $\chi$. For all the terms of this type eq. (\ref{d21}) gives the leading large $N_c$ contribution. However if we consider a term in $F$ containing for example a product of three pairs of $\bar\rho$'s, two of the $\bar\rho$ factors will be  annihilated by the derivatives coming from $\Sigma$. The colour structure in these terms will be different and the action of eq.(\ref{d21}) does not emulate correct large $N_c$ limit for these terms.

We did not find a more simple and compact expression for $\chi^{4d}$ which would reproduce the large $N_c$ limit for an arbitrary term in $F$, than simply 
follow the diagrammatic rules by 't Hooft.
This amounts to the following. The group $SU(N)$ is promoted to $U(N)$. Every adjoint index $a$
is promoted to a pair of a fundamental and an antifundamental indices $a\rightarrow ( ^{\alpha}_{\bar\alpha})$, and every generator $T^a_{bc}$ is represented by an antisymmetric combination of two vertices
\begin{equation}
T^a_{bc}\rightarrow T^{(\alpha\bar\alpha)}_{(\beta\bar\beta)(\gamma\bar\gamma)}\,=\,
\frac{1}{\sqrt2}\,\left[ \delta^{\alpha \bar\beta}\delta^{\bar\alpha\gamma}\delta^{\beta\bar\gamma}\,-\,
\delta^{\alpha \bar\gamma}\delta^{\bar\alpha\beta}\delta^{\bar\beta\gamma}\right]
\end{equation}
The charge density operator $\rho$ also acquires a pair of fundamental/antifundamental indices 
$\rho^a\rightarrow\rho^{\alpha}_{\bar\alpha}$.
Diagramatically then the kernel can be represented as a sum of the terms of the type (we draw a diagram that corresponds to an arbitrary set of four transverse coordinates in the functional derivatives - in the actual kernel those coordinates become $x$, $y$ or $z$ depending on which term in eq.(\ref{b1}) one is representing)
\begin{figure}[htbp]
\begin{center}
\epsfig{file=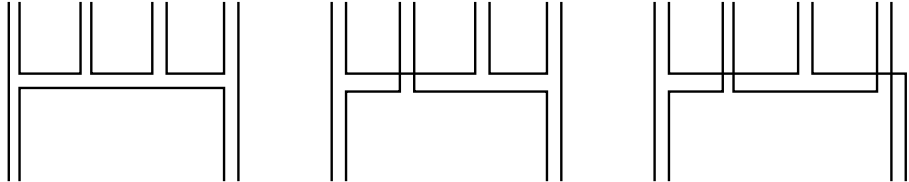,width=120mm}
\end{center}
\caption{\it Double line representation of different terms in the kernel $\chi^{4d}$.}
\label{fig1}
\end{figure}
To see how this produces the leading large $N_c$ result for eq.(\ref{b2}) we also have to represent diagramatically $F$ and $W$. 
The contracted pairs of the charge densities are simply represented (Fig. \ref{fig2}) 
as double wedges (the angle of the wedge is where the indices of the two charge densities are contracted) 
\begin{figure}[htbp]
\begin{center}
\epsfig{file=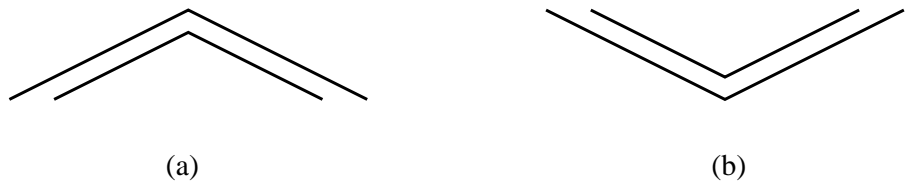,width=120mm}
\end{center}
\caption{\it Double wedge representation of  (a) the projectile charge density, (b) the target charge density.}
\label{fig2}
\end{figure}
Finally to discuss $W$ 
 we take it in the form eq.(\ref{ft}) with $\Sigma$ having a similar structure to $F$, that is where functional derivates come in pairs with pairwise contracted color indices.
Diagrammatically each such pair in $W$ are therefore represented as in Fig. \ref{fig2},b.

With these graphical elements one can calculate the large $N_c$ limit for any particular term in $F$ and $W$. For example taking two pairs of $\rho$ in $F$ and one pair of functional derivatives in $W$ we have the diagrams (Fig. \ref{fig3})
which reproduce the  action of eq.(\ref{d21}).
\begin{figure}[htbp]
\begin{center}
\epsfig{file=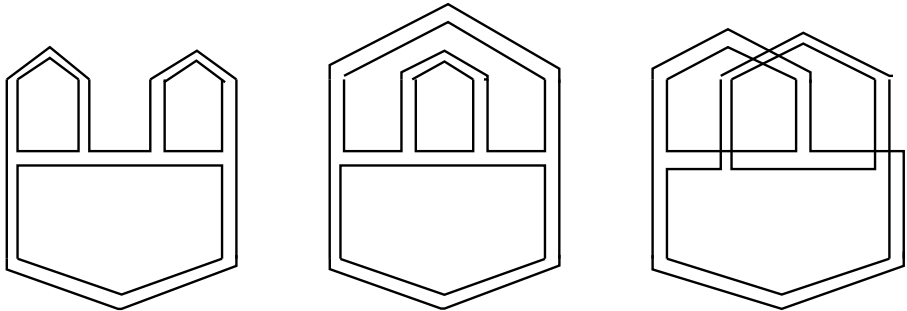,width=120mm}
\end{center}
\caption{\it The planar diagrams contributing to the large $N_c$ limit of the quartic term in $F$. These diagramms are resummed in eq.(\ref{d21}).}
\label{fig3}
\end{figure}
\begin{figure}[htbp]
\begin{center}
\epsfig{file=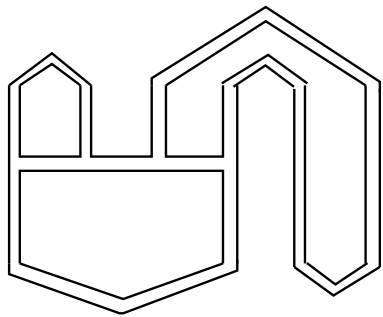,width=50mm}
\end{center}
\caption{\it An example of a planar diagram which contributes to the large $N_c$ limit of the sixth order term in $F$. This type of diagramms are not contained in eq.(\ref{d21}).}
\label{fig4}
\end{figure}
Other terms can be diagramatically represented, and as usual the leading large $N_c$ contribution comes from planar diagrams. To illustrate the fact noted in Section 2 that eq. (\ref{d21}) does not resumm all the leading $1/N_c$ terms we give and explicit diagrammatic example in (Fig. \ref{fig4}).
In Fig. \ref{fig4} the kernel $\chi^{4d}$ is taken to act on six factors of $\rho$ (3 pairs) out of which
four are annihilated by the kernel itself while the remaining ones are removed by the action
of two derivatives in $\Sigma$ of target. Diagramms of this type are not contained in eq.(\ref{d21}).

\acknowledgments
We thank Urs Wiedemann for many discussions on subjects directly relevant to this work. 
We are also grateful to I. Balitsky, J. Bartels, Yu. Kovchegov, L.McLerran, A. Mueller, A. Stasto, H. Weigert and S. Wong for very informative and inspiring discussions during the workshop on Classical and Quantum Apects of Color Glass Condensate, BNL, March 2005.


\begin{thebibliography}{99}
\bibitem{footnote3} The notation KLWMIJ (pronounced: Claw Midge) is not an acronym, but rather the result of DDD transformation 
applied to JIMWLK.
\bibitem{footnotea}The ordering variable $t$ in eq.(\ref{ft}) can be given the meaning of the light cone coordinate $x^+$, as it enters the proper eikonal factor of the target gluon propagating through the hadron $H$. This is however only a philosophical matter, and has no bearing on any of the physics or mathematics discussed in this paper.
\bibitem{footnote} We note in passing, that perturbative gauge invariance does not require the wave function of the target to be globally color invariant. Instead the requirement is somewhat milder, and only refers to the density matrix of the target. Thus in perturbation theory one is allowed for example to have a single gluon as a target on condition that the scattering amplitude is averaged over the color indices, see e.g. \cite{urs}. For these targets (or equivalently projectiles) the argument of \cite{oderon} is incorrect and one cannot replace $K$ by $M$. This is however immaterial in the framework of the dipole model.
\bibitem{footnoteb}We stress that the result of the derivative expansion is unique. In particular we 
could have started with a different representation of the kernel which is obtained using the relation (\ref{h41}), e.g.
$$
\chi^{KLWMIJ}[\rho]\,=\,\frac{\alpha_s}{2\,\pi^2}
\int_{x,y,z}\,K(x,y,z)\,\rho^a(x,\,0)\,\rho^b(y,\,0)\left[1\,+\,
R^\dagger(x)\,R(y)\,-\,R(x)^\dagger R(z)\,-\,R(z)^\dagger R(y)\right]_{ab}
$$
We would still get the same four derivative term (\ref{d2}).
\end{thebibliography}
\end{document}